# Large Domain Graphene

Xuesong Li[1], Carl W. Magnuson[1], Archana Venugopal[2], Eric M. Vogel[2], Rodney S. Ruoff[1], Luigi Colombo[3]

[1]Department of Mechanical Engineering and the Texas Materials Institute, 1 University Station C2200, The University of Texas at Austin, Austin, TX 78712-0292

[2]Dept. of Electrical Engineering, The University of Texas at Dallas, Richardson, TX 75080, USA

[3]Texas Instruments Incorporated, Dallas, TX 75243

Corresponding author e-mail: colombo@ti.com; r.ruoff@mail.utexas.edu

Graphene growth by chemical vapor deposition has received a lot of attention recently *(1)* owing to the ease with which large area films can be grown, but growth of large domain or equivalently large grain size has not been reported yet. In our earlier work we found that growth of graphene on Cu by CVD occurs predominantly by surface nucleation followed by two dimensional growth, but the domain size was limited to a few tens of microns.*(2)* The presence of domain boundaries has been found to be detrimental to the transport properties; the precise mechanism for the degradation still remains elusive but what is known is that these structural defects promote surface reactions with the ambient such as adsorbates or with deposited dielectrics *(3)*. The presence of heptagons and pentagons in the network of hexagons has been observed experimentally *(4)*, and first principle quantum transport calculations predict that the periodicity breaking disorder can adversely affect transport properties *(5)*. Any of these defects can give rise to higher surface chemical activity that would further disrupt the $sp^2$ bonding nature of graphene, and thus impact graphene's fundamental properties.

In this brevia, we report on a CVD process that yields graphene with domains about a factor of 30 larger than previously reported, by very low pressure CVD, less than 50 mTorr, and very low precursor flow rates using methane as the source of carbon on the inside of copper foil enclosures at high temperature, around 1000 ºC. The basic growth process is similar to that previously reported but the growth takes place at much lower methane flow rates yielding a low nucleus density. Fig. 1A shows a scanning electron microscope image of a very large domain with high "edge roughness"; the high "edge roughness" leading to a reactive growth front because of its high surface energy. The shape of the graphene nuclei in the initial stages of growth is also shown in Fig. 1B where a hexagonal symmetry is clearly observed. Fig. 1D further shows the inter-domain structure as the graphene domains grow towards each other to fully cover the surface of the substrate.

The quality of the large area domain films was also evaluated by measuring the transport properties of the graphene films transferred onto silicon dioxide grown on Si wafers. Field effect transistors (FET) were fabricated using Ni metal for the source and drain contacts and the highly doped Si substrate was used as the back-gate contact. The resistance was measured at room temperature as a function of back-gate voltage and the mobility was extracted using the methodology introduced by Kim et al. *(6)*. The mobility for these large domain films was found to be from about 4,000 to 21,000 cm$^2$/Vs, which is within the range of mobility found for exfoliated graphene samples, suggesting that having a large domain minimizes scattering introduced by defects generated at domain boundaries. Raman spectroscopy shows the presence of graphene only, Fig. 3, where the defect D-, the G-, and the 2D-bands are clearly shown. The Raman spectra shown are for 2 regions, one within the film and the other within the dendrite. The G-band full width at half maximum (FWHM) is about the same, 23 cm$^{-1}$, for the two regions and the intensity ratio of the 2D band to the G-band suggests that the carrier concentration is different for the 2 regions*(7)*.

## Acknowledgments

The authors appreciate support from the Nanoelectronic Research Initiative – SouthWest Academy of Nanoelectronics (NRI-SWAN).

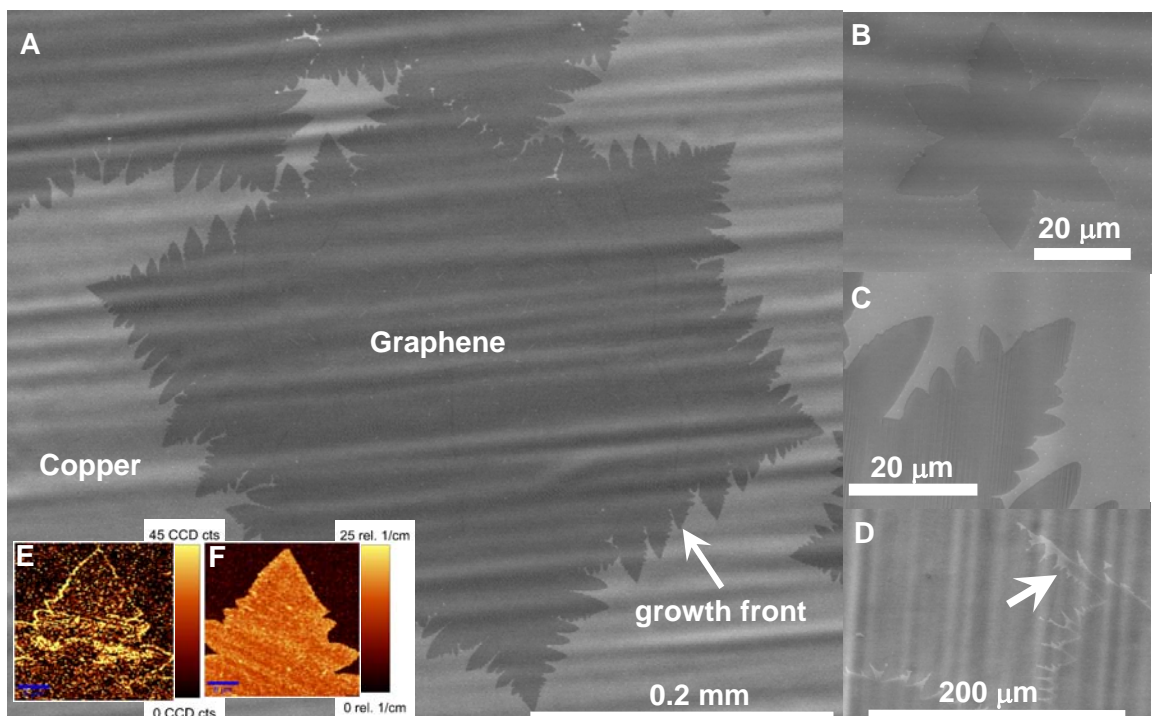

Figure 1. Scanning electron microscopy image of graphene on copper grown by CVD: (A) graphene domain grown at 1035 ºC on Cu at an average growth rate of about 6 µm/min; (B) graphene nuclei formed during the initial stage of growth; (C) high surface energy graphene growth front shown by arrow in (A); (D) graphene domains as they close onto each other (arrow); (E) Raman D-band map of graphene transferred onto sapphire showing a low defect density and defects along the graphene edges, and (F) Raman G-band map the presence of a monolayer of graphite.

**Materials and Methods**

Large domain graphene was grown by chemical vapor deposition (CVD) on copper at very low pressures, ~ 0.04 mbar. We found out that by creating a protective enclosure made of Cu foil we were able to grow extremely large area graphene domains on the inside surface of the Cu foil. The domains first start growing as six-sided polygons and eventually grow into very large graphene domains with growing edges that look like dendrites. Fig. 1 shows the average domain branch length (about half the domain size) as a function of growth time for 2 flow rates under isobaric conditions. The purpose of using an enclosure was to create a more controlled environment during growth. This implementation of CVD graphene led to large domain graphene on the inside of the enclosure and smaller domains but fully covered graphene on the outside with few layer graphene regions dispersed throughout the film. Figure 2 also shows a Raman map of the G-band (Fig 2a) and D-band (Fig 2b) and Raman spectra at different regions of the domain, inside the bulk of the film and within the growing dendrite. The spectra show that the growing material is indeed graphene with a low D-band intensity across the domain.

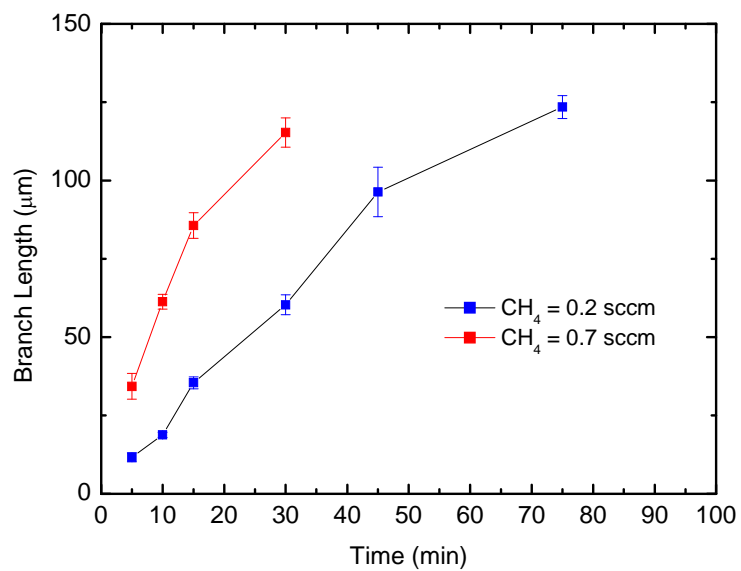

Figure 1. Graphene growth inside the enclosure for two methane flow rates as a function of time under isobaric conditions, $P_{CH4}$ = 30mTorr.

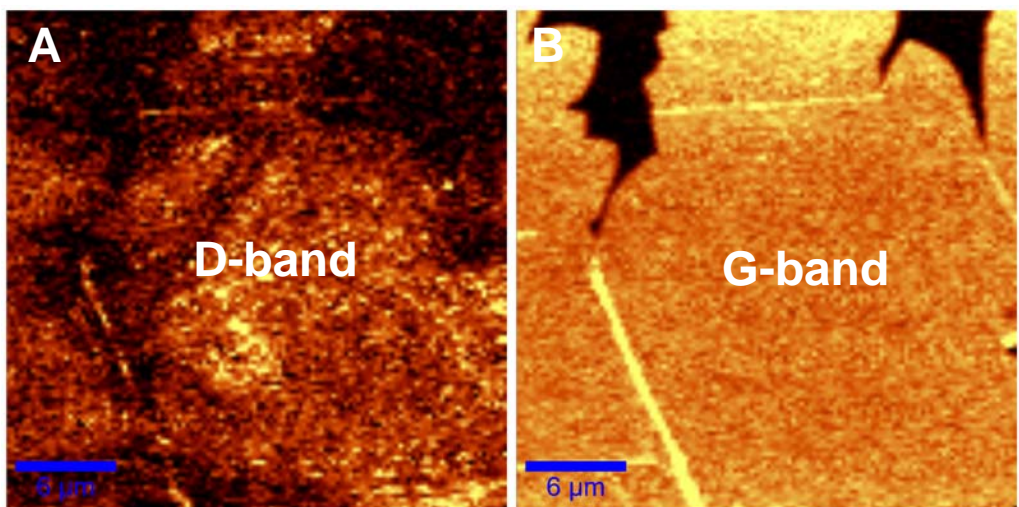

Figure 2. A) Raman D-band and B) G-band map of graphene within the domain and at the edges of the growing domain.

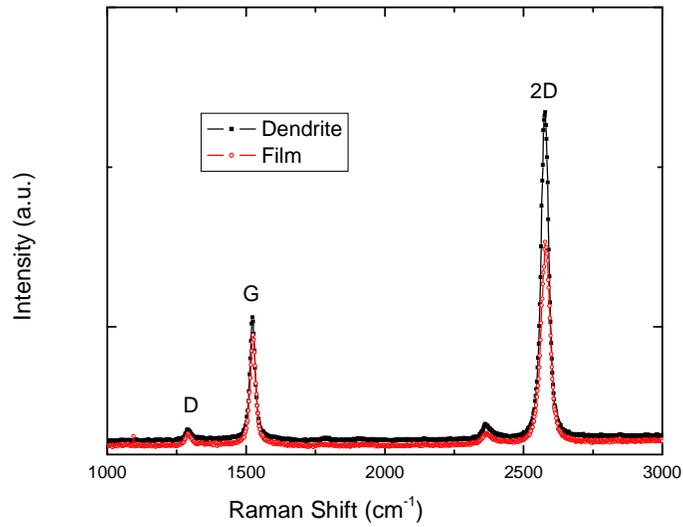

Figure 3. Raman spectra of large domain graphene within the bulk of the film and along the dendrite. The FWHM of the 2D band of the dendrite is slightly lower than that of the bulk and the ratio of the intensity of the 2D band to the G is higher for the dendritic region than the bulk suggesting a lower carrier concentration for the dendritic region. within the "bulk" of the domain and the tip of the dendrite. 2D peak FWHM = 38 (bulk), and FWHM = 32cm$^{-1}$ for the "suspended" dendrite, whereas the FWHM is about 23 cm$^{-1}$ for both regions.